\documentclass[a4paper,11pt]{article}
\usepackage{aaskaiid}
\usepackage{orcidlink}
\usepackage{wrapfig}
\usepackage{subcaption}

\title{The Large-Scale Structure of the Universe through the SKA lenses }
\ShortTitle{Large-Scale Structure}

\author[1,2]{V. Cuciti\orcidlink{0000-0003-4454-132X}}
\ShortName{Cuciti et al.} 
\author[3*]{Surajit Paul\orcidlink{0000-0003-4046-6959}}
\author[4]{Viral Parekh\orcidlink{0000-0001-6282-6025}}
\author[1,2]{Franco Vazza\orcidlink{0000-0002-2821-7928)}}
\author[5]{M. Br{\"u}ggen\orcidlink{0000-0002-3369-7735}}
\author[6]{Prateek Gupta\orcidlink{0000-0002-3716-3618}}
\author[7,8]{Sameer Salunkhe \orcidlink{0000-0002-0666-2326}}
\author[9]{S. Sankhyayan\orcidlink{0000-0003-2601-2707}}
\author[1,2]{A. Bonafede\orcidlink{0000-0002-5068-4581}}
\author[10]{T. Akahori}
\author[2]{G. Brunetti}
\author[2]{R. Cassano}
\author[6]{M. Hoeft}
\author[8]{R. Kale}

\affiliation[1]{Physics \& Astronomy Department, University of Bologna, via P. Gobetti 93/2, I-40129 Bologna, Italy}

\affiliation[2]{INAF - Istituto di Radioastronomia,
              via P. Gobetti 101, Bologna, Italy}  
\affiliation[3]{Manipal Centre for Natural Sciences, Manipal Academy of Higher Education, Karnataka, Manipal 576104, India}
\emailAdd{surajit.paul@manipal.edu}
\affiliation[4]{National Radio Astronomy Observatory, 1011 Lopezville Rd, Socorro, NM 87801, USA}
\affiliation[5] {Hamburger Sternwarte, University of Hamburg,
              Gojenbergsweg 112, D-21029, Hamburg, Germany}
        
\affiliation[6]{Th\"uringer Landessternwarte, Sternwarte 5, 07778, Tautenburg, Germany}

\affiliation[7]{School of Computing, MIT Art, Design and Technology University, Pune, 412201, India}

\affiliation[8]{National Centre for Radio Astrophysics, Tata Institute of Fundamental Research, S. P. Pune University Campus, Ganeshkhind, Pune 411007, India}

\affiliation[9]{Tartu Observatory, University of Tartu, Observatooriumi~1, 61602 T\~oravere, Estonia}

\affiliation[10]{Mizusawa VLBI Observatory, National Astronomical Observatory of Japan, 2-21-1 Mitaka, Tokyo 181-8588, Japan}

\DeclareUnicodeCharacter{207B}{\textsuperscript{-}}

\usepackage[utf8]{inputenc}
\usepackage{newunicodechar}
\newunicodechar{≈}{\approx}

\abstract{

The large-scale distribution of galaxies in the Universe forms an intricate, interconnected network known as the cosmic web. Cosmological simulations within the standard $\Lambda-$CDM framework successfully reproduce this filamentary structure and predict that the nodes and filaments are filled with tenuous plasma at temperatures ranging from $10^5$–$10^8$~K. The hottest and luminous plasma in the nodes corresponds to the intra-cluster medium, while the cooler, more tenuous, gas extends along filaments and cluster outskirts. Galaxies and galaxy groups form and flow along these filaments before accreting onto galaxy clusters (the nodes), outlining the dynamical evolution of large-scale structures. During this process, enormous amount of energy is dissipated through complex plasma processes that can be traced by radio emitting electrons. Despite strong theoretical support for this picture, observational validation remains limited. While massive clusters have been widely detected across various wavelengths, cluster outskirts and the diffuse intergalactic medium within filaments has remained elusive due to their extremely faint emission. The advent of highly sensitive radio facilities such as LOFAR, uGMRT, and MeerKAT has recently enabled a few successful detections of emission from comparatively denser regions of the cosmic-web. These include radio megahalos, permeating the entire cluster volume, as well as bridges of radio emission connecting cluster pairs. 
In this chapter, we summarize current theoretical insights into the cosmic web, discuss observational strategies and recent discoveries, and highlight how the forthcoming Square Kilometre Array (SKA) is expected to transform our understanding of the cosmic web and the distribution of baryons in the Universe.}

\begin{document}
\maketitle

\section{Introduction}
In the standard $\Lambda$-cold dark matter ($\Lambda$CDM) cosmological paradigm, the large-scale structure (LSS) in the Universe emerges through the continuous accretion of matter and the hierarchical merging of smaller structures, ultimately leading to the formation of gravitationally bound structures \citep{Springel_2006Natur}. This process gives rise to a complex network of matter distribution in the late Universe, commonly referred to as the cosmic web. The cosmic web is composed of massive galaxy clusters connected by filamentary structures, interspersed with vast voids and intermediate two-dimensional sheets. 

The state-of-the-art cosmological simulations, typically including both hydrodynamics and dark matter solvers within the $\Lambda$CDM framework \citep{Dolag2008SSRv,Li2008ApJS}, have yielded some of the most detailed visual and quantitative dynamical representations of the cosmic evolution to date \citep{Kolchin2009MNRAS,Nelson2019ComAC}. In these simulations, galaxy clusters, the most massive gravitationally bound systems, form at the intersections of matter filaments.. These nodes represent the densest regions of the cosmic web, where gas and dark matter flow along filaments to fuel the continued growth of these massive structures \citep[]{Kolchin2009MNRAS}. When multiple massive nodes come closer, they even form superclusters \citep{Bagchi2017ApJ,Einasto2019A&A}. The intergalactic medium in the cosmic filaments is found to be not so hot or dense ($10^5 < T < 10^7$~K and $\sim10^{-5}$ particles per cm$^3$) like the intracluster medium (ICM, which is $10^7 < T \sim 10^8$~K and $\sim10^{-3}$ particles per cm$^3$) and is termed as the Warm Hot Intergalactic Medium \citep[WHIM;][]{Dave_2001ApJ}. The over dense regions are dominated by dark matter (DM) followed by luminous baryonic gas, or the intergalactic medium that is eventually forced to fall into the deep potential of these DM channels.

While optical surveys have successfully detected the underlying dark matter distribution of the Universe, tracked by galaxies, the vast reservoirs of intergalactic and intra-cluster medium-which makes up the dominant fraction of baryonic matter in the cosmic web \citep{Lovisari_2021Univ} can only be traced through diffuse emissions such as X-rays, Sunyaev–Zel’dovich (SZ) signals, or synchrotron radio radiation. Unlike optical surveys, diffuse X-ray and radio emissions, which primarily depend on the dynamical activity of large-scale structures \citep{Feretti2012A&ARv}, can also reveal their evolutionary stages and involved thermal and non-thermal plasma physics.

The dynamical evolution of the LSS leads to frequent mergers and interactions, releasing enormous amounts of energy. In particular, the energy released during merger between massive galaxy clusters is of the order of $10^{64}$ erg. This energy dissipates through complex plasma processes, heating the ICM and surrounding regions, and generating shocks as well as hydro and magneto-hydrodynamic (MHD) instabilities \citep[e.g.][]{1999ApJ...518..594R, 2000ApJ...535..586T,Paul2011ApJ}. The hot ICM emits X-rays primarily via thermal bremsstrahlung, whereas shocks and MHD turbulence accelerate charged particles \citep{Bykov_2008SSRv,Brunetti2011MNRAS} and amplify cluster-scale magnetic fields \citep{Beresnyak2016ApJ}. The relativistic particles subsequently lose energy through synchrotron radio emission \citep{Feretti2012A&ARv, Brunetti2014IJMPD}. Spectacular evidence of these non-thermal phenomena in galaxy clusters are Mpc-sized diffuse radio sources, such as radio halos and radio relics. Radio halos are centrally located sources, typically following the X-ray emission of the cluster \citep{Cassano01.2026.SKA}. Radio relics are found at the periphery of clusters and have elongated shapes \citep{ArpanPal01.2026.SKA}. Radio halos are powered by turbulence, injected in the ICM during merger events, whereas particles in radio relics are accelerated by merger driven shock waves \citep{Brunetti2014IJMPD,Weeren_2019SSRv}. These diffuse radio sources represent unique laboratories for investigating the mechanisms of energy dissipation, particle acceleration and magnetic field amplification during the formation of the LSS of the Universe.

Beyond testing these predictions of $\Lambda$-CDM cosmology, deep radio observations of galaxy clusters and the cosmic web will probe the distribution of matter on cosmic scales, measure the growth and geometry of structure and critically confront both the $\Lambda$-CDM paradigm and compelling non-standard alternatives. For an example, modified gravity and dynamical dark energy scenarios predict a scale- and redshift-dependent growth rate, leading to measurable signatures in redshift-space distortions and the HI 21 cm power spectrum \citep{Maartens2015,Bull2015}. More relevant to this chapter, a significant fraction of baryons inferred from early-Universe measurements remains undetected at low redshift. As discussed earlier, cosmological simulations with standard $\Lambda$-CDM suggest that most of these baryons reside in the warm–hot intergalactic medium (WHIM) within filaments, though direct confirmation remains limited \citep{Shull2012}, leaving room for alternative interpretations. Alternative dark matter models, such as warm or self-interacting dark matter, suppress small-scale clustering and alter filament thickness, halo abundance, and void statistics \citep{Bode2001,Vogelsberger2012}. Primordial non-Gaussianity would introduce a characteristic scale-dependent bias on ultra-large scales \citep{Camera2015}, while deviations from statistical isotropy could manifest as anomalous radio dipoles in wide-area continuum surveys \citep{Bengaly2018}. Collectively, these observables enable critically test the fundamental assumptions underlying mainstream and alternative cosmological models. 
 
While galaxy clusters are routinely detected in X-rays \citep{Bulbul_2024A&A}, as well as through SZ \citep{Planck_collab2016A&A} and radio observations \citep[see][for reviews]{Weeren_2019SSRv, Paul_2023JApA}, their outskirts, the filamentary connections between clusters, and supercluster environments remain largely unexplored. This difficulty primarily arises due to the low surface brightness (both radio and X-ray) of emission in these regions that are believed to host the warm–hot intergalactic medium (WHIM), characterized by exceedingly low densities and temperatures.

The advent of new-generation low-frequency radio telescope array such as LOFAR, MeerKAT, and the upgraded GMRT—offering unprecedented sensitivity—has however opened up the possibility of detecting extremely faint, diffuse sources on scales extending beyond individual galaxy clusters. Low frequency observations have the unique capabilities of tracing extremely aged plasma and extremely inefficient phenomena, such as those common in galaxy clusters and outside \citep{deGasperin01.2026.SKA}. LOFAR has recently discovered that some clusters host Megahalos, diffuse sources filling the volume of clusters up to their periphery \citep{cuciti22}. In a few cases, LOFAR detected large radio structures connecting pre-merging clusters, called radio bridges \citep{govoni19, botteon20}. The direct observations of radio emission from cosmic filaments has not been achieved yet, but stacking experiments reveal evidence of excess of radio emission in overdense regions of the cosmic web \citep{vernstrom21}. These discoveries represent the first steps in the exploration of the LSS of the Universe in the radio band. 

In the future, the SKA, with its proposed unparalleled sensitivity and survey speed, is expected to discover thousands of galaxy clusters in a new regime of mass and redshift which revolutionize the cluster abundance studies \citep{Cassano2015,Maartens2015, Cassano01.2026.SKA}. In synergy with upcoming large galaxy-redshift, X-ray and Sunyaev–Zeldovich (SZ) surveys, the galaxy cluster observation with SKA provides a powerful cosmological tool to well constrain the cosmological parameters, e.g., matter density $\omega_{m}$, Hubble constant $h$, and matter fluctuation amplitude $\sigma_{8}$ \citep{Allen2011}. This capability enables SKA to break degeneracies between competing models and tightly constrain early-Universe physics \citep{Bull2015}.

In this chapter, we first review the current theoretical understanding of the LSS (Section \ref{sec:theory}). We then discuss recent discoveries of extended radio emission beyond the usual galaxy cluster halo (Section \ref{sec:observations}). Finally, we outline how radio galaxies—within clusters, superclusters, and even voids—can inform us about the evolution of the LSS (Section~\ref{sec: BCG}). In each section, we also highlight the prospects for future SKA observations in the AA4 era.

\section{Theoretical understanding of large-scale structures in the Universe}
\label{sec:theory}
The formation of large-scale structures in the Universe is a fundamental prediction of cosmological theories and broadly corroborates with the observational findings \citep{Sahni1995}.
Within the standard $\Lambda$CDM framework, the cosmic web arises from the gravitational amplification of small primordial density perturbations in an otherwise nearly homogeneous early Universe. Observations of the Cosmic Microwave Background (CMB) revealed a primordial fluctuations of order $\sim10^{-5}$ and exhibited nearly Gaussian statistics, consistent with predictions from inflationary models of the early Universe \citep{Planck2020,Komatsu2011,Baumann2009}.

Since in the $\Lambda$CDM paradigm the non-relativistic (“cold”) dark matter dominates the matter density in the Universe, it collapses first under gravity into overdense regions, giving rise to a web-like network of dark matter haloes and filaments \citep{White1978,Springel2005}. In this scenario, baryonic matter, subsequently falls into these dark-matter potential wells, cools radiatively, and forms stars, galaxies and groups that trace the underlying dark-matter distribution in the Universe \citep{Blumenthal1984,Mo2010}.

On sufficiently large scales, the evolution of density perturbations is well described by linear perturbation theory \citep{Peebles1980}, whereas at smaller scales, non-linear collapse through gravitational instabilities, mergers, and the emergence of the cosmic-web morphology become crucial for understanding galaxy, group, and cluster formation \citep{bond1996Natur,Vogelsberger2020}.
While the Zel’dovich approximation provides a first-order analytical description of this anisotropic collapse, numerical simulations becomes essential for modelling the full non-linear evolution of dark matter haloes, filaments and cosmic-web in the Universe \citep{Asgari2023}.

\subsection{Cosmological simulations of LSS}

In the era of ever-increasing computational power and advances in numerical methods and statistical tools, it is now feasible to simulate large fractions of the Universe in a finite, expanding volume \citep{springel01, Kolchin2009MNRAS, 2002ApJ...566..103N}. In the history of cosmological simulations, the first significant milestone is generally considered to be the N-body integration by \cite{Press1974ApJ}, which modeled the mass distribution of bound clumps formed through hierarchical clustering.
Over the last few decades, the development of faster numerical methods and the advent of increasingly powerful processors have enabled a large number of extensive cosmological simulations, e.g., Millennium \citep{Springel2005}, Millennium-II \citep{Kolchin2009MNRAS}, Bolshoi \citep{Klypin2011ApJ}, Illustris, and IllustrisTNG \citep{Vogelsberger2014MNRAS, Nelson2018MNRAS, Springel2018MNRAS}.

\begin{figure}[h!]
	\centering
	\begin{subfigure}{0.49\textwidth}
		\includegraphics[width=\linewidth]{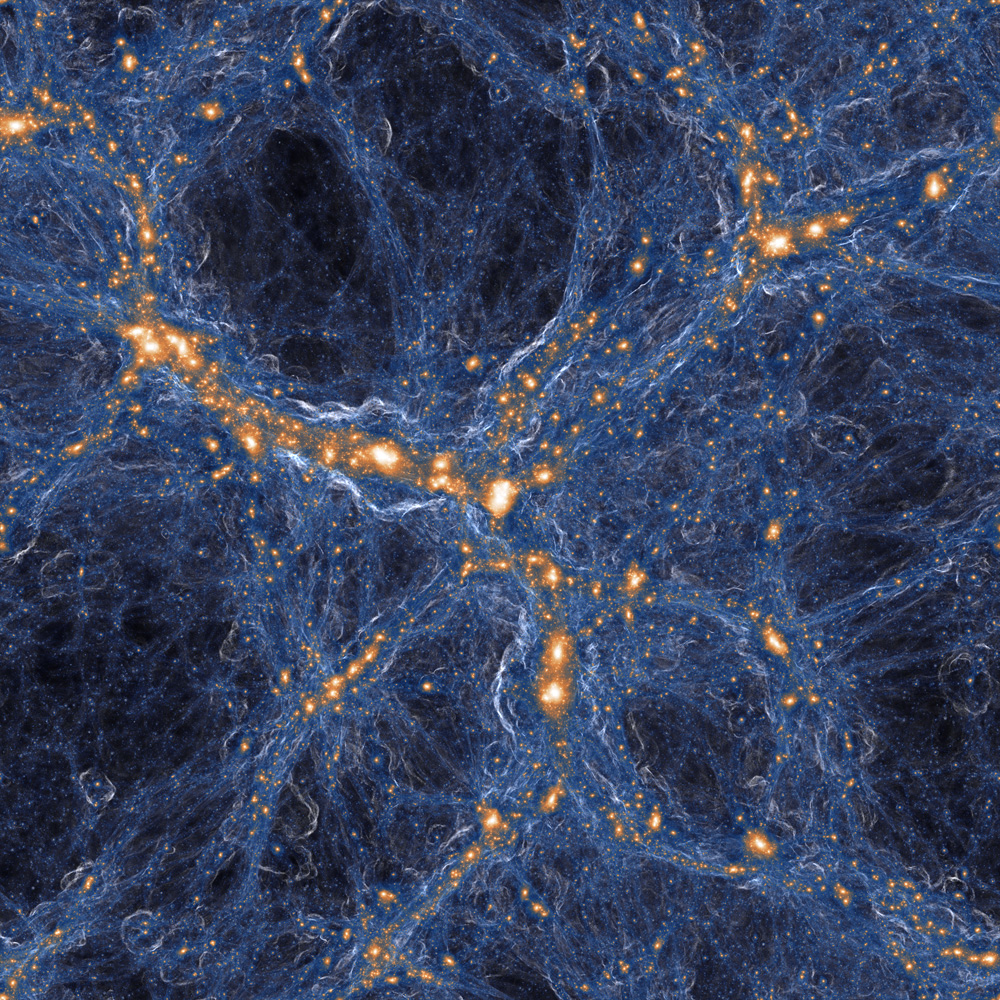}
	\end{subfigure}
	\begin{subfigure}{0.49\textwidth}
	        \includegraphics[width=\linewidth]{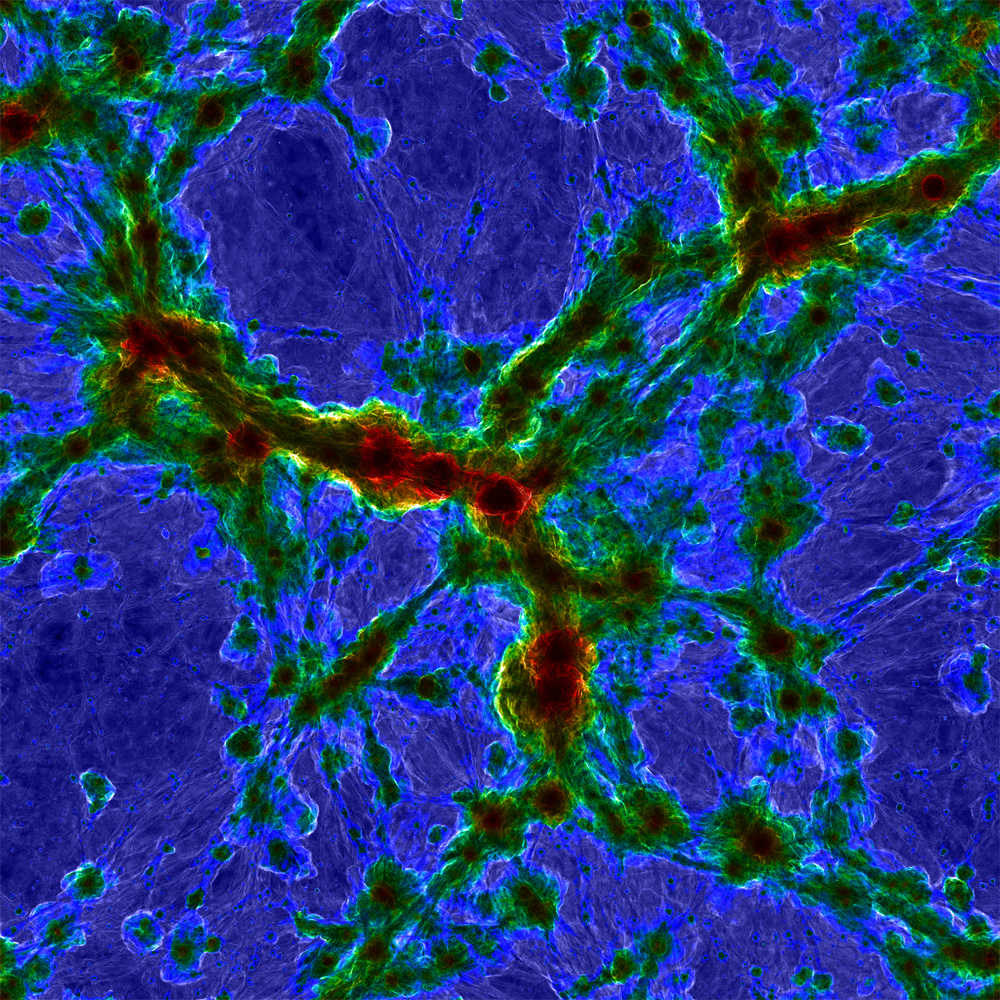}
         \end{subfigure}
	\caption{Left panel: A projection map of dark-matter density overlaid with the average Mach number of shocks along the line of sight. All the gravitationally-collapsed structures (shown in orange/white) are surrounded by successive shock surfaces (in blue), which encode their formation histories.
Right panel: A projection map combining gas temperature (colour) and shock Mach number (brightness). Red indicates ~10 million Kelvin gas at the centres of massive galaxy clusters; the bright outer structures trace diffuse intergalactic-medium gas being shock-heated at the boundaries between cosmic voids and filaments. Image credit: IllustrisTNG Collaboration.}
	\label{fig:ShockinTNG100}
\end{figure}

The Millennium Simulation \citep{Springel2005} was the first true cosmological scale simulation. The simulation was performed using \textsc{GADGET} \citep{Springel2005MNRAS} cosmological code that followed the growth of structures using $2160^3$ dark-matter particles within a cubic volume of approximately 2.23 billion light-years on a side. 
This simulation provided the first robust testbed for comparing theoretical predictions with observational data, revealing remarkable agreement between the geometric properties of simulated superclusters and those observed in the 2-degree Field Galaxy Redshift Survey (2dFGRS; \citealt{Colless2001MNRAS, Colless2003astro.ph}; see also \citealt{Einasto07, Einasto2007A&AIII}). Adopting a $\Lambda$CDM cosmology, the Millennium Simulation demonstrated that dark matter evolves from an initially smooth, nearly uniform distribution into a highly clustered state \citep{Springel_2006Natur}, with galaxies forming along these dark matter structures. 

The recent IllustrisTNG \citep{Nelson2018MNRAS, Pillepich2018MNRASb, Springel2018MNRAS, Nelson2019ComAC} simulations are comparatively smaller in size than the Millennium simulation, however, include more physics, especially baryonic physics and magnetohydrodynamics that helps visualizing the luminous features of the LSS. Tracing the evolution of baryonic particles in IllustrisTNG facilitates its use as an ideal simulation for studying gas distribution at the outskirts of the galaxy clusters (see Figure~\ref{fig:ShockinTNG100}), and the dynamics of the warm hot intergalatic medium (WHIM) in the filaments, the part of the Universe that is yet to be explored observationally. Transient events associated with structure formation, such as mergers, drive non-thermal processes including shocks and turbulence at cluster peripheries and along filamentary boundaries (Figure~\ref{fig:ShockinTNG100}). These phenomena and the particle acceleration mechanisms that they trigger are discussed in more detail in Section~\ref{sec:non-thermal}.

\subsection{Identification of cosmic-web and its components through algorithms}

The visual appearance of the LSS of the Universe reveals a tapestry of dense and underdense regions. Yet, the quantitative identification of these structures, extending well beyond megaparsec scales, remains a major observational challenge. The main bottleneck lies in the lack of sufficiently deep all-sky surveys across the relevant wavebands. X-ray observations currently offer the most direct probe of the tenuous intergalactic medium (IGM; \citealt{bregman2007, nicastro2017}). However, an all-sky X-ray survey with the required depth is not feasible with existing instruments. Even the sensitivity of eROSITA \citep{eRosita_main, predehl2021}, the deepest all-sky X-ray survey to date, falls short of detecting the extremely faint emission from cluster outskirts and the tenuous filaments of the cosmic web \citep{eckert2015, ursino2016, tuominen2022}. Moreover, achieving surveys significantly deeper than eROSITA is unlikely in the near future.

Radio observations face similar limitations. The expected synchrotron surface brightness from cluster outskirts and cosmic-web filaments is exceedingly low \citep{vernstrom2017, govoni19}. Moreover, as this diffuse emission typically exhibits a steep radio spectrum, they are expected to be detected only at low-frequency observations \citep{Vazza2015A&A, Brunetti2014IJMPD}. Incidentally, only a few arrays currently operate at sufficiently low frequencies and with wide sky coverage, such as LOFAR, uGMRT, MWA, and LWA. However, even the deepest LOFAR surveys, including LoTSS and targeted deep fields, have yielded only tentative detections or statistical hints of emission beyond galaxy clusters that we discuss in the section~\ref{sec:observations}.

These observational constraints underscore that detecting cosmic-web structures requires targeted, extremely sensitive observations. This, in turn, highlights the importance of robust, algorithmic identification of the cosmic web and its components using existing all-sky optical surveys. Structure-finding algorithms can reconstruct the LSS by predicting the locations of massive nodes that may host radio mega-halos, the geometry and orientation of filaments and superclusters, and their physical conditions through the analysis of all-sky survey data aided by cosmological simulations (as elaborated in subsequent sections). The resulting catalogues of promising targets enable the optimization of deep, pointed observations with current sensitive radio facilities and, in the near future, with SKA \citep[e.g.][]{Sousbie2011MNRASa, Libeskind2018MNRAS, oakpaul2023}. By directing observations toward the most promising regions of the cosmic web, these methods substantially enhance the likelihood of detecting its extremely faint radio signatures.

\subsubsection{Identification of structures from simulated mock catalogues}\label{section Halo finder} 
Cosmological simulations trace the time evolution of a representative patch of the Universe. Snapshots saved at different redshifts are analysed to probe the underlying theoretical framework. Each snapshot contains information on how matter and energy densities evolve under the influence of gravitational forces arising from the matter distribution. This evolving distribution naturally gives rise to the theoretically expected components of the LSS of the Universe, especially those beyond the galaxy clusters. Because these components exhibit distinct physical properties, dynamics, and geometries, a wide range of structure-finding algorithms has been developed over the past few decades \citep{bond1996Natur, Cautun2014MNRAS,Libeskind2018MNRAS} to quantitatively characterize the cosmic web and to aid in predicting its observable signatures with current and future facilities. Among these structures, dark matter halos (observationally, galaxy clusters) and filaments (the connective framework of the cosmic web) are of particular interest, as they lie at the interface between observational data and theoretical predictions \citep{Knebe2011MNRAS, Tempel2014MNRAS}.

Dark matter halos are the densest regions on large scales and are typically identified as locally overdense, gravitationally bound systems. Algorithms developed for this purpose, collectively known as halo finders, are widely employed in cosmological analyses. Most halo finders \citep{Knebe2011MNRAS, Onions2012MNRAS, Knebe2013MNRAS, Gupta2023PRD} are based on either the spherical overdensity (SO) method \citep{Press1974ApJ} or the Friends-of-Friends (FoF) algorithm \citep{Davis1985ApJ}, with many extensions combining features of both approaches. More advanced methods, such as the ROCKSTAR algorithm \citep{Behroozi2013ApJ}, use full six-dimensional phase space and temporal information to robustly identify substructures and tidal features. The recently developed MITRO algorithm \citep{Gupta2023PRD} instead applies mass overdensity criterion to isolate actual gravitationally bound structures.

In contrast, filament identification is more challenging. The average dark matter density in filaments is roughly two orders of magnitude lower than in halos at the nodes of the cosmic web \citep{Cautun2014MNRAS}, and their complex, elongated morphologies further complicate automated detection. Consequently, most filament-finding methods rely on geometrical or morphological criteria \citep[see][and references therein]{Libeskind2018MNRAS}. Recently, machine-learning–based classifiers have also been employed to characterize filamentary structures \citep{Buncher2020MNRAS, Inoue2022MNRAS, Awad2023MNRAS}. Using the DisPerSE filament finder \citep{Daniela2024A&A}, a comprehensive catalogue of cosmic filaments has been constructed from the MillenniumTNG simulation, enabling detailed studies of their global properties across cosmic time. These theoretical advances in mapping and characterizing the LSS strongly motivate deeper, multiwavelength observational studies, which will not only help validate and constrain cosmological models but also refine our understanding of underlying astrophysical processes.

\subsubsection{Identification of structures from large scale surveys}\label{algo-structure}

At present, the most detailed and deepest all-sky surveys are available at optical wavelengths (e.g. SDSS, Pan-STARRS, Gaia; \citealt{ahumada2020, chambers2016, gaia2023}). However, optical observations primarily yield data on individual galaxies, cannot directly image the diffuse intergalactic medium and, therefore, offer little information about the baryonic content distributed within the cosmic web \citep{Shull2012}. Nevertheless, the wide-area coverage and precise redshift measurements make SDSS extremely valuable for identifying suitable regions in three dimensions, thereby enabling targeted follow-up observations at wavelengths relevant to the detection of diffuse material and the WHIM \citep[][and Devika P. C. et al., in prep.]{oakpaul2023}. Since optical wavelengths predominantly trace stellar emission rather than the diffuse ICM, the identification of galaxy clusters from the optical redshift surveys is necessarily indirect and often computationally demanding. Typically, this is accomplished using clustering or group-finding algorithms, most commonly based on the FoF or SO methods 
as described in section~\ref{section Halo finder}.

\begin{wrapfigure}{r}{0.5\textwidth}

\vspace{-0.3cm}\includegraphics[width=0.99\linewidth]{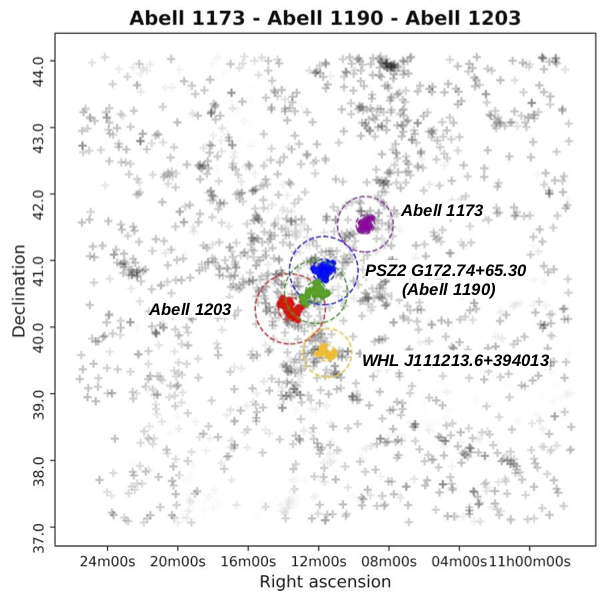}
    \caption{SDSS galaxies overplotted by identified interacting clusters \citep{oakpaul2023}.}
    \label{fig:sdss_oak}
    \vspace{-0.3cm}
\end{wrapfigure}

Numerous studies have sought to identify and characterize galaxy clusters using such optical datasets over the past decades \citep[e.g.,][]{Yoon_2008ApJS, Tempel_2012A&A}. However, the early systematic attempts to identify merging clusters, extending beyond simple cluster detection—were made by \citet{Tempel_2017A&A}. More recently, \citet{oakpaul2023} presented a detailed analysis of the dynamical states of interacting cluster systems, identifying massive nodes and probable filamentary connections within the cosmic web (see Fig.~\ref{fig:sdss_oak}). The authors also reported indications of diffuse emission (especially in radio waves) associated with selected cosmic nodes, as seen in survey maps. A second catalogue of similar findings has been reported very recently by \cite{Wen_2024MNRAS}. The extensive list of nodes identified in these two studies offers a valuable resource for future deep searches for diffuse emissions, particularly in the context of detecting the diffuse intergalactic medium within the cosmic web—an effort poised to benefit greatly from the projected capabilities of SKA in future.

Simulations indicate that, besides clusters near nodes, rare configurations exist where chains of massive clusters form wall-like superclusters. The Shapley Supercluster, identified by Somak Raychaudhury in 1989 \citep{Raychaudhury1989}, was the first major example, and many superclusters have since been catalogued \citep[e.g.][]{Einasto1996_supercluster_void_network, Martinez14, Sankhyayan2023, Liu2024}. These studies have revealed extensive networks of interacting cosmic-web nodes and numerous superclusters 
\cite{Sankhyayan2023}, highlighting regions of concentrated luminous baryonic matter \citep{Cautun2014MNRAS, Dolag2006} through comparison with cosmological simulations. Moreover, simulations predict that shocks, turbulence, and accretion flows associated with structure formation in these regions will produce low-level diffuse radio emission \citep{Vazza2015A&A, vernstrom2017, Brown2017}.

Observational efforts targeting known superclusters have already begun to uncover tentative diffuse radio emission in inter-cluster filaments and bridges, both in LOFAR surveys \citep{Duchesne2022, Botteon2022} and in targeted observations connecting massive clusters \citep{govoni19}, demonstrating that algorithmically identified supercluster regions are ideal for deep, low-frequency radio observations with current facilities and the upcoming SKA (see Section~\ref{sec: BCG}). Consequently, robust supercluster catalogs not only map the LSS but also provide critical guidance for targeted searches of the faint diffuse radio signatures of the cosmic web.

\subsection{Predictions for diffuse radio emission from the components of cosmic web}
\label{sec:non-thermal}

The growth of cosmic structures is intertwined with the release of shock waves, at whose surface the kinetic energy of infalling gas gets dissipated into gas thermalisation \citep[e.g.][]{1972A&A....20..189S,2003ApJ...593..599R,Bykov_2008SSRv}. Around all most massive structures in the cosmic web, i.e. from clusters of galaxies to the filaments connecting them, strong $\mathcal{M} \geq 10-10^2$ accretion shocks (where $\mathcal{M}=v_s/c_s$ is the Mach number, $v_s$ is the shock velocity and $c_s$ is the local sound speed) should exist over cosmological timescales, and mark the approximate transition from smooth rarefying Universe and the overdense cosmic regions undergoing the virialisation process.

A couple of decades of cosmological numerical simulations has established a picture in which the number statistics of shocks in the Universe is predicted to be dominated by structure formation shocks driven by gravity \citep[e.g.][]{mi00, 2003ApJ...593..599R,Vazza2011MNRAS}, with only a minor contribution from non-gravitational processes \citep[e.g.][]{ka07,2016MNRAS.461.}. On the other hand, the bulk of the dissipation of infall kinetic energy happens in the $2 \leq \mathcal{M} \leq 4$ range, at overdensities in the range of matter halos. Such shocks are often associated with accretions and mergers within massive halos, and are supposed to be a main contributor of the enrichment of the intracluster/intragroup medium with cosmic rays \citep[e.g.][]{pfrommer06,John2019MNRAS}.

From the theoretical viewpoint, however, the regime of weak ($\mathcal{M} \leq 5$) shocks is also the one at which making predictions of Diffusive Shock Acceleration (DSA, \citealt[e.g.][]{1996A&A...309.1002O,2001RPPh...64..429M,2013ApJ...764...95K}) gets more difficult, owing to a number of physical and numerical uncertainties, in particular related to the uncertain ``micro''-details of shock acceleration models \citep[e.g.][and discussion therein]{2019SSRv..215...14B}. In addition to shocks, the acceleration and re-acceleration of cosmic ray electrons, protons and heavier nuclei by large-scale plasma motions stirred by the same matter accretion events have been proposed to be provided by Fermi-type acceleration \citep[e.g][and references therein]{Brunetti2014IJMPD}.

Finally, the amplification of magnetic fields is also predicted to follow the growth of cosmic structures, via the small-scale dynamo amplification \citep[e.g.][]{1999A&A...348..351D,2004ApJ...612..276S,2006MNRAS.366.1437S,donnert18}, although the reality is likely to be more complex than the ideal MHD picture usually explored via cosmological simulations or theory \citep[e.g.][]{2016PNAS..113.3950R}. This amplification process in the highly turbulent plasmas is deemed to be sufficient to explain the typically observed $\sim \rm \mu G$ magnetisation levels in clusters of galaxies \citep[e.g.][]{Beresnyak2016ApJ}, while its efficiency is probably much less in cluster outskirts and in cosmic filaments, owing to the shorter available turbulent forcing timescales and to the mostly super-sonic nature of turbulence there \citep[e.g.][]{2005ApJ...631L..21B,2014MNRAS.445.3706V}.

While $\mu$G magnetic fields have been successfully measured in galaxy clusters \citep[e.g.][]{gf02,2003ApJ...597..870E,vacca10,bonafede10,2013MNRAS.433.3208B}, only limits in the order of $\leq 0.075-0.25 \rm ~\mu G$ were derived for nearby large cosmic filaments using LOFAR \citep[e.g.][]{locatelli21,2023MNRAS.523.6320H}, while the stacking detections by \citet{vernstrom21} and \citet{vernstrom23} derived tentative values in the range $\sim 30-60 \rm ~nG$ for the magnetisation of the radio emitting fraction of filaments (see Sec.\ref{subsec:shocks}). The average radio spectrum of emission inferred from the latter observations is also compatible with a simple $I(\nu) \propto \nu^{-1}$ spectrum, consistent with the theoretical predictions from DSA for the acceleration of relativistic electrons by high Mach number shocks \citep[e.g.][]{keshet04}.

Therefore, the exciting possibility of directly detecting the radio signal from strong shocks in filaments with upcoming radio facilities will be crucial: it could be used to inform us about the level of magnetisation in regions where a dynamical memory of magnetic seed fields should still be present, which ultimately can lead to understand the origin of cosmic magnetism \citep[e.g.][]{vazza21,2025Univ...11..164C}, which is presently up for several possible generation mechanisms \citep[e.g.][]{1997PhRvL..78.3610B,2016RPPh...79g6901S,2021RPPh...84g4901V}. Based on modern numerical simulations of the evolution of primordial magnetic fields, jointly with the additional release of magnetic fields from galaxies, in Section \ref{sec:simulations} we present the updated analysis for the realistic chances of detecting the radio signature of structure formation shocks from the cosmic web with the upcoming SKA telescopes.

\section{Observations of components of cosmic-web in radio and the role of SKA in future}
\label{sec:observations}

Although SKA is expected to revolutionise our knowledge of the non-thermal components through the cosmic web, big steps forward have been taken in the past few years with SKA precursors and pathfinders. In this Section we discuss these recent descoveries, some of the open points and the expectations for SKA observations derived from cosmological simulations. 

\subsection{Beyond galaxy clusters: radio Megahalos}
\label{Sec:megahalos}

\begin{figure}[ht]
    \centering
	\includegraphics[width=\textwidth]{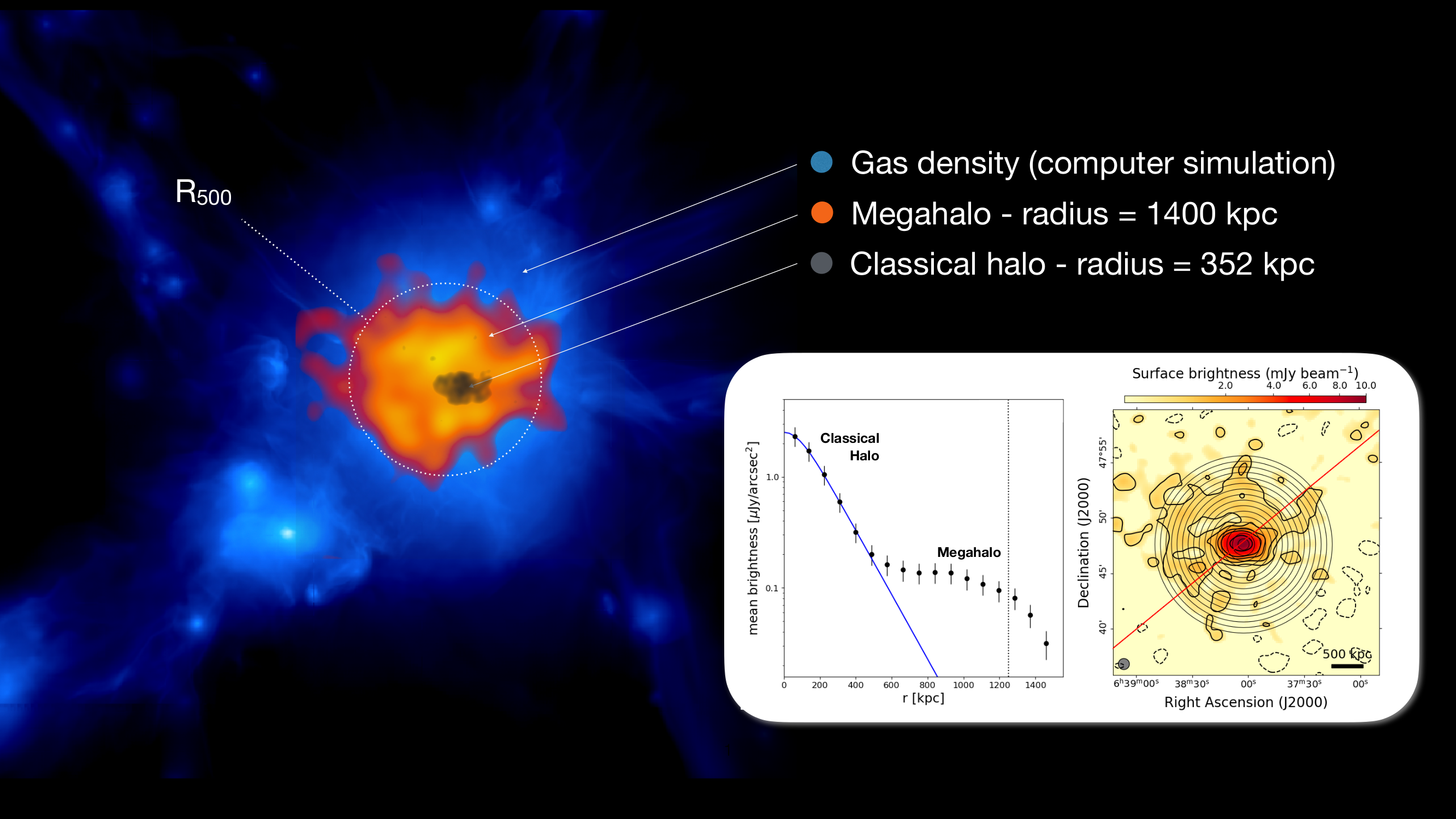}
    \caption{Blue: simulated gas density distribution \citep{vazza19}. Orange: LOFAR image of the megahalo in Zwcl 0634.1+4750. Grey: LOFAR image of the radio halo in Zwcl 0634.1+4750. Inset: Surface brightness profile of the radio emission in Zwcl 0634.1+4750 \citep{cuciti22}.}
    \label{fig:megahalo}
\end{figure}

Using LOFAR observations, \citet{cuciti22} found four clusters where the radio halo is embedded in an emission that is more than 30 times larger in volume but $20-30$ times lower in emissivity and has a shallower surface brightness profile than radio halos (Fig.~\ref{fig:megahalo}). This sources, named ``Megahalos'', fill the volume of the clusters, up to at least $R_{500}$ \footnote{$R_{500}$ is the radius within which the mean mass over-density of the cluster is 500 times the cosmic critical density at the cluster redshift. The mass enclosed in a sphere with radius $R_{500}$ is $M_{500}$.}, therefore reaching the most peripheral regions of galaxy clusters. 
The surface brightness of the radio halos at the center of these clusters can be well fitted with an exponential function in the form $I = I_0 e^{-r/r_e}$, as it is usually done for radio halos \citep[][]{balboni24}. The surface brightness of megahalos instead departs from the exponential function and remains fairly constant over a wide range of distances from the clusters’ centre (Fig.~\ref{fig:megahalo}). X-ray observations of galaxy clusters show that over the same range of radii, the thermal gas density usually goes down by a factor of $\sim$few \citep{ghirardini19}, which means that the non-thermal components become energetically relevant in the external region of clusters. This demonstrate that megahalos can give us important constraints about the characteristics of the plasma in clusters' outskirts.
So far, spectral information is only available for two megahalos and it is based on marginal detections with LOFAR at 54 MHz. Constraining the spectrum of diffuse sources is of paramount importance to understand the acceleration mechanisms that power them \citep[e.g.][]{brunetti08nature}. The available data on these two megahalos suggest a spectrum $\alpha\sim -1.6$, steeper than radio halos ($\alpha\sim -1.3$), indicating that the mechanisms of acceleration must be inefficient. For this reason, turbulence was proposed as a good candidate \citep{nishiwaki24}, but further information about the spectrum of megahalos is needed.

Another illustrative case is that of the massive cluster \textit{PLCK\,G287.0+32.9}. Using uGMRT observations, \citet{salunkhe25} reported diffuse radio emission extending over $\gtrsim 3$~Mpc with two components. The second component has a steep spectrum ($\alpha < -1.5$) and a nearly flat surface-brightness profile, consistent with the characteristics of a radio megahalo, detected for the first time at frequencies higher than those probed by LOFAR. Subsequent deep MeerKAT observations by \citet{rajpurohit25} reported even more extended emission from the cluster. Owing to its higher sensitivity and superior angular resolution at gigahertz frequencies, MeerKAT revealed a greater number of discrete sources and several extended filamentary features within the peripheral region of halo, which may account for the deviation in the surface-brightness profile (giving the second halo component) reported earlier. To address, this point, which may in principle affect all the detected megahalos, we simulated a LOFAR observation of a galxy cluster with a radio halo and added hundreds of faint (2-3 $\sigma$ level) point sources spread over the cluster region (Cianfaglione et al., submitted). We subtracted them from the \textit{uv}-data with usual
procedures and computed the surface brightness profile after the subtraction. The resulting profile does not show any significant deviation from the exponential law, suggesting that in the detected megahalos the second component cannot be entirely explained by the presence of faint sources that have not been properly subtracted, However, the contrasting results in the case of \textit{PLCK\,G287.0+32.9} warn us that extreme caution is needed in distinguishing genuine cluster-scale emission from blended components. In this respect, the combination of SKA-Low, with its superior surface-brightness sensitivity and wide \textit{uv}-coverage, and SKA-Mid, with its higher resolution and sensitivity, will be crucial in disentangling overlapping structures and provide precise spectral mapping to constrain the dominant particle-acceleration processes in cluster outskirts.

The discovery of megahalos triggered the first cosmological simulation aimed at reproducing the different components of diffuse emission in galaxy clusters. Specifically, \citet{beduzzi23} simulated one galaxy cluster following the position of radio emitting particles during the lifetime of the cluster. They showed that the cluster's peripheral regions can be filled with electrons that have been energised by turbulence up to energies sufficient to emit at LOFAR frequencies. The same simulation also suggests that in some cases, depending on the epoch of the simulation and on the observing line of sight, the radio emission produced by those electrons has a surface brightness profile resembling the ones observed in megahalos \citep{beduzzi24}.
\begin{wrapfigure}{l}{0.55\textwidth}
\includegraphics[width=0.55\columnwidth]{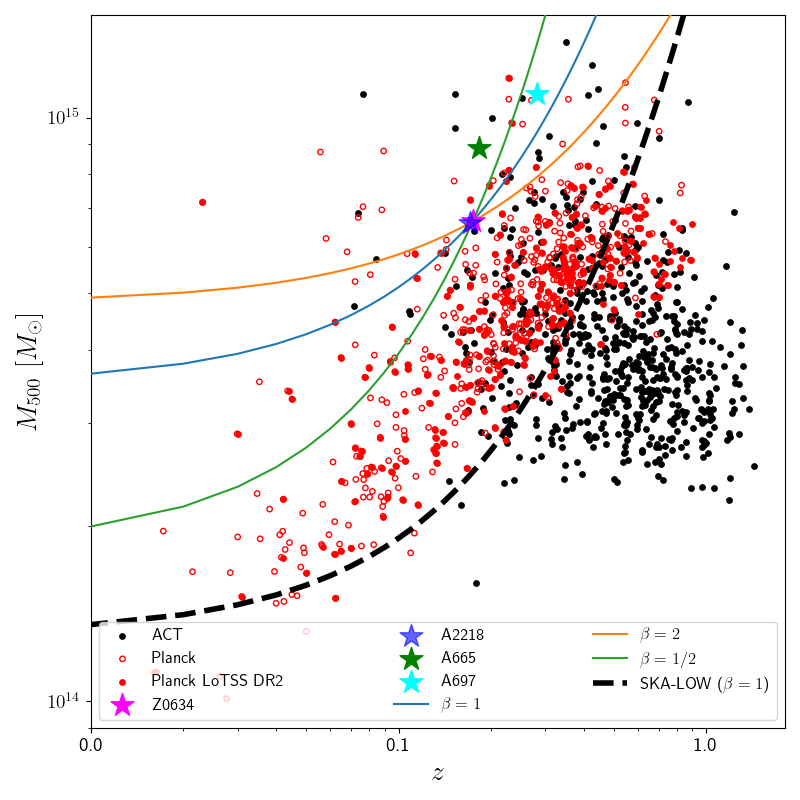}
    \caption{Mass–redshift diagram for all clusters in the Planck sample (red dots). Red filled dots indicate clusters included in the Planck LoTSS DR2 sample, while black dots represent clusters from the ACT catalogue. Stars mark clusters hosting megahalos. Solid lines show the region where megahalos are detectable with LoTSS, and the dashed line indicates the region where they are expected to be detectable with SKA-Low. Adapted from \citet{cuciti22}.
    }
    \vspace{-1cm}
    \label{fig:M-Z}
\end{wrapfigure}
Although still limited to a few cases, the observation of megahalos demonstrates that relativistic electrons and magnetic fields fill the whole volume of clusters, and that we can use radio observations to probe the dynamics of large-scale structures and the dissipation energy mechanisms therein.

\subsubsection{The role of SKA}

Fig.~\ref{fig:M-Z} shows the mass--redshift diagram for the Planck clusters in the second data release of the LOFAR Two Meter Sky Survey \citep[LoTSS,][]{shimwell22} (black dots). Interestingly, megahalos (stars) are all located in the most massive clusters of the sample, suggesting that a proportionality between the megahalo brightness and the cluster mass is at play (as for radio halos). The three solid lines show the expected mass–redshift relation taking into account the cosmological surface brightness (SB) dimming (SB$\propto (1+z)^{-4}$) and assuming a power-law dependence of the large-scale emission surface brightness with the mass of the clusters \citep[SB $\propto M^{\beta}$, where we considered three possibilities for $\beta$,][]{cuciti22}. As the large-scale emission in ZwCl 0634.1+4750 is detected at around 3$\sigma$, we impose the condition that the lines go through that point in the diagram to set the normalization.
With these assumptions, the region of the mass-redshft diagram where we expect to be able to observe megahalos with LoTSS observation is the region on the top left of the solid lines in Fig.~\ref{fig:M-Z}. In fact, the four megahalos discovered in \citet{cuciti22} are all located in this region. \citet{botteon22sci} found emission filling the entire volume of the cluster Abell 2255 and Cianfaglione et al., (in prep.) discovered two new megahalos and two candidate ones, also in massive clusters lying in the expected parameter-space region. This reinforces our idea that we might be seeing just the tip of the iceberg of a widespread phenomenon that could be unveiled with deeper observations. 

Given the steep spectrum of megahalos, SKA-Low is the most suitable instrument to detect them, however, the combination with SKA-Mid will be crucial to address the spectral properties of these sources, needed to disentangle the underlying physical mechanisms, and to identify discrete sources embedded in the diffuse emission. With one hour SKA-Low observation we estimated an rms noise of 11.6 $\mu$Jy/beam with a beam of $9.1''\times7.4''$ (using briggs=0) at a central frequency of 200 MHz. Considering a steep spectrum for megahalo of $-1.6$, this would mean that SKA-low would be $\sim$ 3 times more sensitive than LOFAR HBA for the detection of megahalos (black dashed line in Fig.~\ref{fig:M-Z}). This will give us the possibility to search for megahalos in the vast majority of the clusters of the Planck sample (red dots in Fig.~\ref{fig:M-Z}). We also note, that being SKA in the Southern Sky, we will be able to exploit also the SZ catalogue from the Atacama Cosmology Array (ACT), which is more complete than Planck, especially at high redshift.

\subsection{Accretion and external shocks}
\label{subsec:shocks}

In addition to merger shocks, which are usually found around $R_{500}$, galaxy clusters are expected to be surrounded by accretion shocks, which should be located around $2-3 R_{200}$ ($\sim 5$ Mpc from the center of a massive cluster). With respect to merger-induced shocks, which typically have Mach numbers of the order of a few, the temperature of the upstream intergalactic medium is low compared to the already shock heated gas in the vicinity of galaxy clusters, leading to much stronger shocks with Mach numbers of 10-100.
Accretion shocks are sites of efficient energy conversion where the gravitational energy of infalling material is converted into thermal energy. Therefore, accretion shocks play a significant role in the definition of the dynamics of galaxy clusters and the thermalization of the ICM. Accretion shocks are expected to not only accelerate particles, but they should also align magnetic field, thus producing highly polarised synchrotron radiation. 
\citet{vernstrom23} stacked all-sky radio maps in polarisation at the position of pairs of luminous red galaxies and they detected a stacked polarised signal, with polarisation fractions of $\sim$20\%, which is interpreted as the organisation of local magnetic fields by strong shock waves both at the cluster peripheries and between clusters. These works also allowed the authors to estimate magnetic field strength in filaments to be in the range $30-60$ nG, based on equipartition and IC arguments. Although the assumption of equipartition can be questioned given the little knowledge we have about the plasma conditions in filaments, it shall be noticed that similar values of magnetic fields where also inferred using the complementary approach of testing different magnetic field models dynamically evolved with cosmological MHD simulations  (similar to the ones used in Sec. \ref{sec:simulations}), and assuming conservative efficiencies for the injection efficiency of relativistic electrons by cosmic shock waves \citep[][]{vernstrom23}. The same set of simulations and magnetic field models have been validated against several other radio observations \citep[e.g.][for a recent review]{2025Univ...11..164C} hence the range of magnetic fields inferred by the stacking of filaments is presently well supported by data.
These results are also in line with upper limits on magnetic field strength given by \citet{locatelli21} based on the non-detection of diffuse emission between cluster pairs with LOFAR.
\citet{hou23} stacked the low frequency radio data from OVRO-LWA of 44 X-ray bright clusters and found a significant (4-5$\sigma$) excess around $2.5 R_{500}$. Interestingly, such radio emission is coincident with a previously claimed $\gamma-$ray signal detected in stacked Fermi observations, which was interpreted as inverse-Compton emission from accretion shocks.

\subsection{Bridges and filaments}
\label{sec:bridges}

\begin{figure}[ht]
    \centering
	\includegraphics[width=0.45\columnwidth]{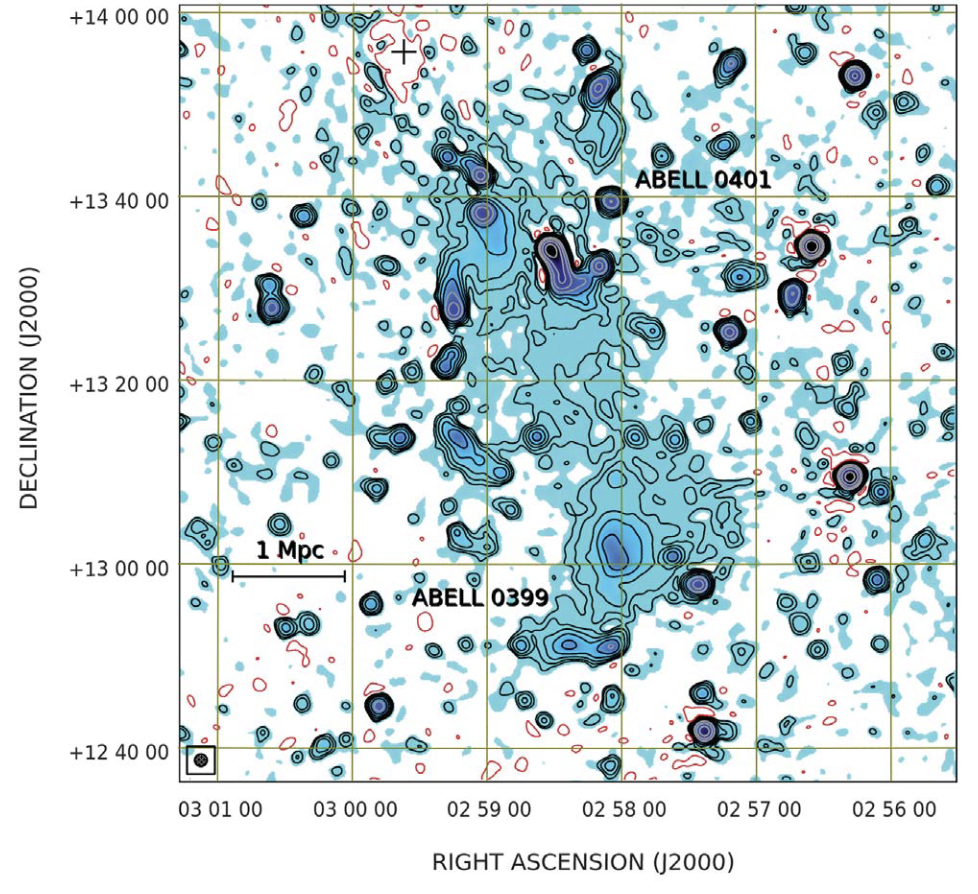}
    \includegraphics[width=0.47\columnwidth]{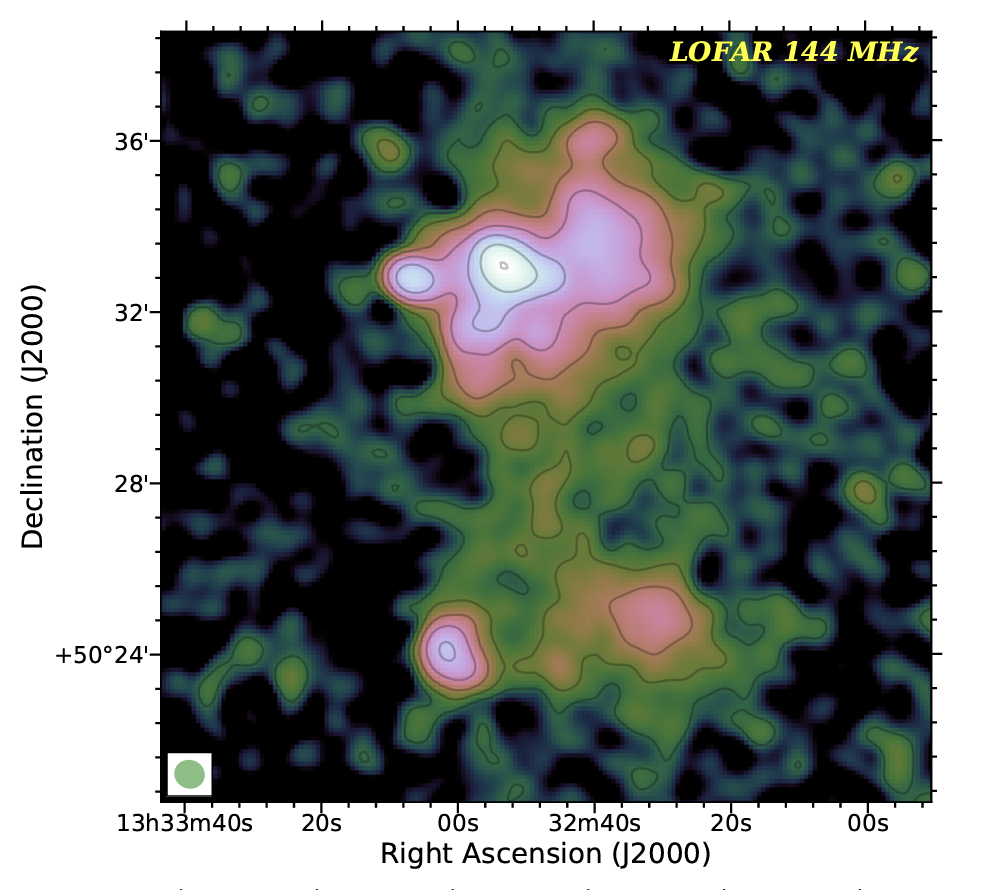}
    \caption{Radio bridges. Left: A399 -- A401 \citep{govoni19}. Color and contours show the radio emission at 144 MHz with a resolution of $80''$ and rms= 1 mJy/beam. Contours start at 3 mJy/beam and increase by factors of two. Right: A1758 \citep{botteon20}. Color and contours show the radio emission at 144 MHz with a resolution of $35''$ and rms= 160 $\mu$Jy/beam. Contours start at 3 $\sigma$ and increase by factors of two. }
    \label{fig:bridges}
\end{figure}

Regions in-between cluster pairs are at the crossroads between the denser and hotter ICM regions and the colder and rarefied cosmic web, thus representing an ideal target to study thermal and non-thermal gas outside galaxy clusters. 
In this respect, LOFAR observations have discovered large-scale radio emission (3 Mpc long, projected) between the pre-merging clusters A399 and A401, called ``radio bridge” \citep[Fig.~\ref{fig:bridges},][]{govoni19}. A similar emission has also been found between the pre-merging clusters 1758N-S \citep[Fig.~\ref{fig:bridges},][]{botteon20}.
The point to point radio X-ray analysis of these bridges reveals a good correlation between the thermal and non-thermal emission \citep{botteon20, deJong22}, suggesting that they are connected and originate from similar volumes.
\citet{pignataro24a} have used deep LOFAR LBA observations and obtained the spectral index distribution in the bridge connecting A399 and A401. Using LOFAR LBA, HBA, and uGMRT observations, they performed a multi-frequency spectral study of the bridge. Between 60 MHz and 144 MHz, they found an integrated spectral index value $\alpha= -1.44 \pm 0.16$. They also found that the radio spectrum shows a significant steepening between 144 and 400 MHz. Such a steepening can be explained in a turbulent re-acceleration framework, assuming that the acceleration timescales are longer than $\sim$200 Myr.
Although based only on a marginal detection of the A1758 bridge at 53 MHz, \citet{botteon20} constrained the spectral index of the bridge to be $\alpha_{53MHz}^{144MHz}=-1.65\pm 0.27$. 

\citet{govoni19} proposed that radio bridges may result from first-order Fermi (Fermi-I) re-acceleration of a volume-filling population of fossil relativistic electrons by weak shocks under favorable projection effects. Alternatively, \citet{brunettivazza20} suggested that the synchrotron emission from radio bridges could be a result of second-order Fermi (Fermi-II) re-acceleration, where turbulence plays a major role in amplifying magnetic fields and re-accelerating particles. This simulation predicts that the emission of bridges should have steep spectra ad should be more volume filling with respect to the case of shock acceleration. Although the statistics is still scarce, the steep spectrum of bridges, combined with the fact that the bridge emission is volume filling, suggests turbulence as the main mechanism for particle acceleration.

More recently, two new candidate radio bridges have been discovered: in the clusters pairs A2061--A2067 \citep{pignataro24b} and A3016--A3017 \citep{hu25}, which however do not fill entirely the space between the clusters. In these two cases, one of the two clusters does not host a radio halo, which might be the reason for the observed ``gap'' in the bridges. However, there is no correlation between the radio and X-ray point to point surface brightness in these systems, therefore the possibility that they are linked to the interaction with a sub-group is still open.

In addition, bridges of radio emission have been found also filling the space between clusters and groups. A famous example is in the Coma cluster, connecting with the NGC4839 group \citep{bonafede22}. Another bridge has been recently found with MeerKAT observations in the Shapley supercluster, connecting A3562 and the group SC1329–313 \citep{venturi22}.

While bridges arise from the dynamical activity that takes place through cosmic filament, they still represent exceptional places in the Universe, where matter is compressed and heated by the upcoming merger between two clusters. The diluted and colder plasma in the filaments of the cosmic web has been elusive so far. Only stacking experiments have successfully found evidence of radio emission from filaments. In particular, \citet{vernstrom21} used multiple all-sky radio and X-ray maps to stack pairs of luminous red galaxies, which are commonly used as tracers for galaxy clusters and groups. In both the X-ray and radio stacked maps they found emission of $>5 \sigma$ significance on scales larger than 3 Mpc.

\subsection{Predictions for SKA}
\label{sec:simulations}
To produce an updated prediction of the fraction of the cosmic web which may become detectable with the SKA, we analysed a recent MHD cosmological simulation from a suite introduced in \citet[][]{va25}. These new MHD simulations were produced with a customised version of the ENZO code \footnote{\url{https://enzo-project.org}} and they improved on previous modelling of the detectability of the cosmic web discussed in \citet{2015aska.confE..97V,Vazza2015A&A,vazza19}, in the sense that these new results include a more realistic view of cosmic magnetic fields, cosmic rays and astrophysical processes. In detail, they:
 a) assume an initial stochastic magnetic field (mimicking primordial generation mechanisms) as primary seed for the magnetic field in cosmic volume. The initial magnetic vectors were drawn from a random realization of a power-law spectrum: $P_B(k) dk \propto k^{-\alpha_B} dk$, while the normalization was set by requiring that the rms field strength,$\langle B^2_{1Mpc}\rangle^{1/2}$, obtained after smoothing over a fixed scale ($\lambda=1 \rm ~Mpc$) is consistent with upper limits from the CMB analysis (e.g. \citealt{2019JCAP...11..028P}). b) The simulations incorporate updated galaxy formation routines for star formation and feedback from AGN, tailored to reproduce the cosmic star formation history, the distribution function of stellar mass in galaxies, the stellar mass fraction in galaxies, the luminosity distribution of radio galaxies, and the correlation between supermassive black hole mass and the host halo gas mass. c) Both star formation and AGN feedback release at run-time additional magnetic fields, corresponding to a fixed ($10\%$) fraction of their energy, mimicking the ubiquitous process of magnetisation by galaxies, which roughly follows the cosmic star formation history; d) These simulations also track at run-time the propagation of a passive cosmic ray fluid, injected both by shocks (identified on-the-fly in the simulation), by AGN and by star formation. The transport of cosmic rays is followed assuming they are perfectly trapped by magnetic fields (which is good enough for $\leq 10~ \rm GeV$ electrons and for the relatively coarse spatial resolution of the simulation) and thus that they perfectly follow baryonic matter. While this model lacks a run-time spectral ageing model for cosmic rays (numerically very expensive), it allows us the approximate estimate of the time since the last injection of cosmic rays in any cell of the simulation, using a technique outlined in \citet{va25}. This allows us  to produce realistic predictions for the full energy spectrum of particles in the simulation.
 In particular, we can compute the expected synchrotron radio emission from the total spectrum of relativistic electrons present in any given cell at any time, which is reconstructed with good approximation by knowing the time elapsed since the last injection of cosmic rays in the same cell, and by applying time-dependent solutions to the spectral energy evolution of the population, based on the local measured values of density, magnetic field and based on the redshift of the observation. This information is extracted from a large pre-computed suite of relativistic electrons spectra  ($\sim 10^6$ model spectra in total), and the corresponding synchrotron emission for different frequencies is assigned to the cells. This approach relies on a prescription, based on the Diffusive Shock Acceleration, to link the efficiency of electron acceleration to the shock Mach number \citep[][]{2024JKAS...57..155K}, resulting at most in a fraction $\approx 10^{-3}$ of the number density of thermal electrons to be injected into supra-thermal electrons for strong shocks.

Among the several plausible stochastic primordial seed fields simulated in the suite, we focus here on the one which best reproduces both the observed trend with redshift of the Residual Rotation Measure\footnote{The Residual Rotation Measure is defined as 
$\rm RRM =RM - GRM$, where $\rm RM$ is the measured RM of an extragalactic source and $\rm GRM$ is the estimated galactic RM component. The latter is of course not known a priori, and different approaches have been proposed to measure it, e.g. by computing the GRM at each source position from the Galactic RM map by \citet{2022A&A...657A..43H}, and taking the median of a 1-degree diameter disc centred at the source. See \citet[][]{2025A&A...693A.208C} for more details and tests on this procedure.} for $z \leq 3$ polarised sources, obtained by \citet{2025A&A...693A.208C} using LOFAR, as well as the positive detection of the stacked radio emission from filaments in the cosmic web, obtained both in total or polarised intensity by \citet{vernstrom21,vernstrom23}. In particular, the favoured magnetic field model is an inflationary-like spectrum with magnetic energy on large scales, $P_B(k) \propto k^{-1}$, with initial field amplitude $\langle B^2_{1Mpc}\rangle^{1/2}\sim 0.37 \rm ~nG$ (comoving), as discussed in \citet{2025Univ...11..164C}.

\begin{figure}[ht]
    \centering
	\includegraphics[width=0.99\columnwidth]{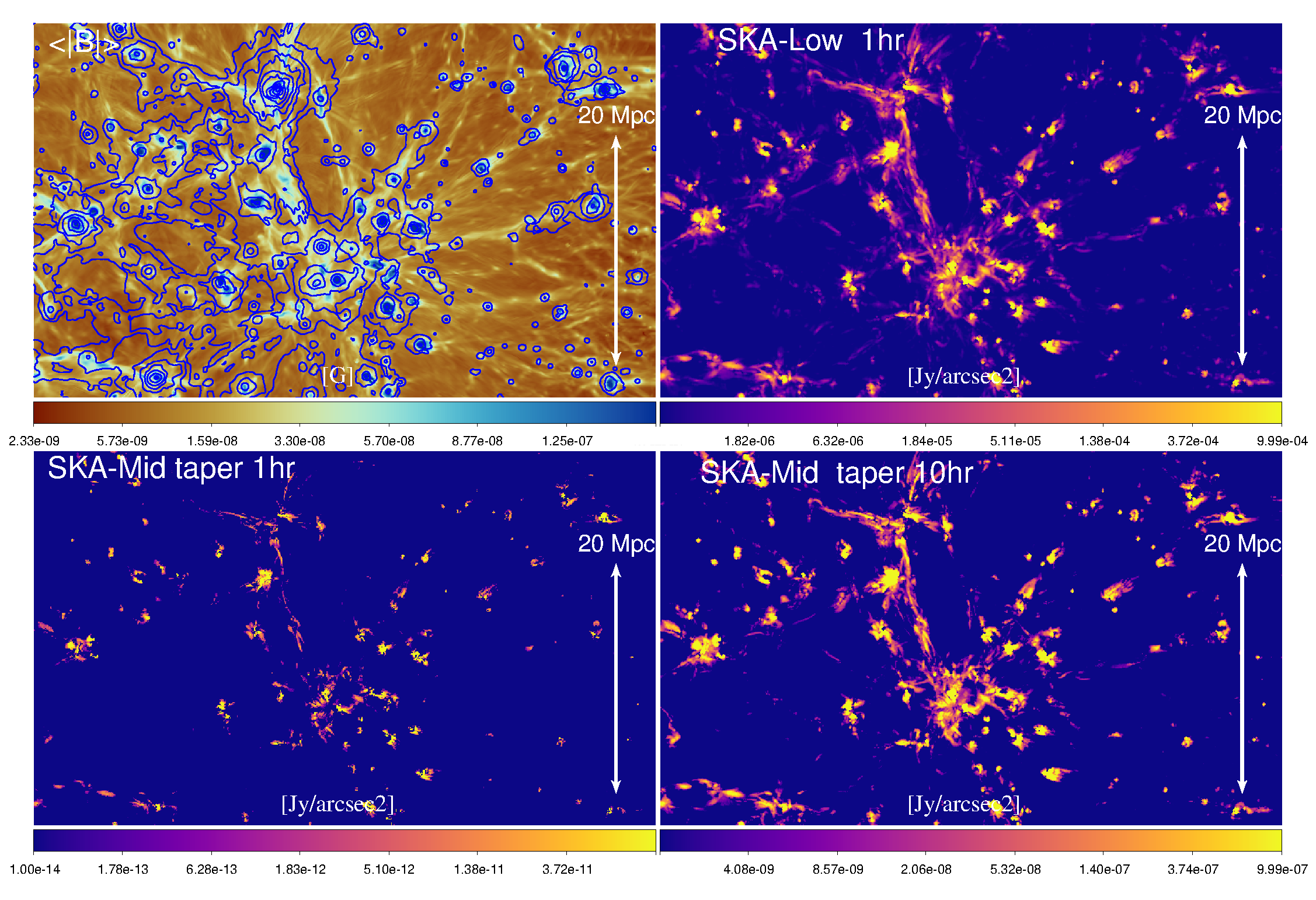}
    \caption{Top left panel: projected (mass weighted) mean magnetic field strength  (colours) and  contours of mass-weighted mean gas temperature along the line of sight of a cosmological simulation. The other panels show the predicted synchrotron radio emission from cosmic shocks, detectable with different observing configurations: a)  with 1hr exposure time with SKA-Low ($50-250$ MHz, rms = 21.2 $\mu$Jy/beam with beam =$13.1''\times10.6''$); b) with 1 hour SKA-Mid band 1 ($0.25-1.05$ GHz, rms = 55.1 $\mu$Jy/beam with beam =$34.5''\times 32.3''$); c) with 10 hours SKA-Mid band 1 (rms =$3.2 \mu$Jy/beam, with beam =$34.5''\times 32.3''$). Only the emission above $3\sigma_{rms}$ is plotted in both cases.}
    \label{fig:cosmo_sim}
\end{figure}

Here we focus on the simulation of this initial field within a comoving $85^3 \rm Mpc^3$ volume simulated with $1024^3$ uniform cells with ENZO, which we analyse in a snapshot at $z=0.15$, which is similar to the average redshift of filaments analysed by \citet{vernstrom21,vernstrom23}. 
Fig. \ref{fig:cosmo_sim} shows the projected radio emission from a fraction of the projected volume in the simulation, which by design has been chosen to focus on the most overdense structures in the simulation, i.e. a large association of clusters of galaxies with total masses of a few $\sim 10^{14} M_{\odot}$, connected by filaments and threads of matter with a projected distance of $\sim 10 \rm ~Mpc$. This crowded cosmic volume gives an idea of possibly the most promising situation where to look for very extended emission beyond the virial radius of clusters of galaxies, and in massive cosmic filaments. 
We explored two frequency bands: 
a) at 200 MHz and observed with SKA-Low ($50-350$ MHz), b) at 800 MHz and observed with SKA-Mid band 1 ($0.35-1.05$ GHz). In detail, we estimated the sensitivity of the SKA-Low band 1 in AA4 configuration assuming 1 hour integration time, with Briggs weighting scheme and Robust parameter equal to 1 (to slightly decrease the resolution). We obtained rms = 21.2 $\mu$Jy/beam with beam =$13.1''\times10.6''$, corresponding to $0.13 \mu$Jy/arcsec$^2$ (Top right panel in Fig. \ref{fig:cosmo_sim}). Such a noise level is close to the confusion limit for the SKA-Low. For the SKA-Mid band 1 observation, we assumed 1 hour integration time in AA4 configuration, Briggs weighting with Robust parameter equal to zero and a taper of $28.1''$. We obtained a confusion limited rms noise of 55.1 $\mu$Jy/beam with beam =$34.5''\times 32.3''$, corresponding to $0.043 \mu$Jy/arcsec$^2$ (Bottom left panel in Fig. \ref{fig:cosmo_sim}). To increase the sensitivity to the diffuse emission, we estimated the noise that we could achieve with 10 hours observation with SKA-Mid band 1, considering that we will use the full resolution, untapered image (estimated noise rms=$0.7 \mu$Jy/beam, with beam=$1.33''\times 1.14''$) to subtract point sources and then we will taper down to $28.1''$, achieving an rms noise of $3.2 \mu$Jy/beam, corresponding to $0.0025 \mu$Jy/arcsec$^2$ (Bottom right panel in Fig. \ref{fig:cosmo_sim}). In the simulated images, we smoothed the angular resolution of the simulation at this redshift to roughly correspond to the one of the SKA-Low observation, while the angular resolution of the simulation is nearly equal to the SKA-Mid one considered here.  The synchrotron emission originated in filaments, where the magnetic fields are compressed by shocks, is predicted to be polarized. On one hand, polarization observations are much less limited by confusion noise due to the lower density of polarized radio sources with respect to total intensity ones, thus allowing one to perform long exposures without hitting confusion. On the other hand, even in the most polarised regions, the polarised intensity would be a fraction of the total intensity shown in Figure \ref{fig:cosmo_sim}. In this respect, an extensive discussion on the possibilities of detecting polarised emission in clusters and filaments with SKA observations is presented in \citet{Vacca01.2026.SKA}.\\
Consistent with earlier modelling \citep[][]{2015aska.confE..97V,Vazza2015A&A}, the simulation shows that SKA-Low is overall better suited to detect the faint and $\sim \rm Mpc$ wide emission from the shocked cosmic web in total intensity,
for a 1 hour integration time. SKA-Low is better suited at sampling larger spatial scales compared to SKA-Mid, and it can also better probe the distribution of low energy electrons which can accumulate even in the outskirts of clusters of galaxies, owing to the integrated effect of ejection from galaxies and previous injection by shocks \citep[e.g.][]{va25}, and which makes the spectral slope of the emission slightly steeper than the canonical $\alpha \approx -1 $ spectrum expected from strong shocks ($\mathcal{M} \gg 1$). However, the 10 hour integration with SKA-Mid with subtracted point sources seems to have the potential of reaching the same detection as the 1 hr SKA-Low case, opening the possibility to perform spectral studies of these diffuse sources. 
The detection of accretion shocks surrounding filaments and clusters of galaxies remains extremely challenging, owing to their low particle density and magnetic fields; however in a crowded region like this, the detection of extremely extended and flat spectrum emission associated with shocks in filaments appear feasible, both in SKA-Low and SKA-Mid (with a ten times longer exposure). 
It has to be noticed that this simulation does not include the additional contribution from turbulent re-acceleration, which can further illuminate close intracluster bridge configurations, as discussed in Section \ref{sec:non-thermal}.

\section{Probing the cosmic web with radio galaxies}
Galaxies are the fundamental building blocks of the LSS of the Universe, tracing out the filamentary network of the cosmic web. They also provide a powerful, multi-scale view to probe this web, from the dense nodes of superclusters to the most under-dense void regions. The spatial distribution of galaxies within superclusters helps to map how baryons and dark matter are arranged across large-scale structures. Comparing their radio properties across supercluster environments allows us to link galaxy evolution to the underlying cosmic web.
\label{sec: BCG}

\subsection{BCGs and Supercluster core}
\par Brightest cluster galaxies (BCGs) are the most luminous and massive galaxies, often found at the core of the galaxy cluster \citep{2015MNRAS.453.1201H,2014MNRAS.440..588H,2007ApJ...663L..85B}. BCGs have a broad band of emission from ultraviolet to far infrared and also H$_{\alpha}$ lines, indicating the presence of multiphase gas and ongoing star formation activities. BCGs are also detected in radio wavelengths and many times their radio emission goes well beyond the optical envelope, in the form of extended radio jets. These powerful radio BCGs (radio-loud) are associated with AGN and exhibit periodic radio jet activities that affect the surrounding ambient medium. These radio BCGs or radio AGNs play an important role in controlling the general growth and evolution of the core part of the cluster \citep{2012ARA&A..50..455F,2012NJPh...14e5023M,2018NatAs...2..198H,2018NatAs...2..179C}. They regulate the cooling flow mechanism, star formation activities, and thermal gas properties (temperature, entropy, and pressure) near the center of the clusters. Studying the nuclear activity of these special galaxies in relation to the wider cluster environment is therefore important. This has important implications not only within the ICM dynamics and BCG evolution but also for comprehending AGN feedback processes globally. Radio BCGs trace local gas pressure, turbulence, and cluster–cluster interactions, and helping to map the physical conditions of galaxy clusters.

\par Galaxy clusters do not exist in isolation but are embedded in the LSS, assembling into superclusters that trace the densest nodes of the cosmic web. Because BCGs reside at the centers of massive clusters, they naturally lie in these high-density environments and can probe the cores of the surrounding LSS. The core part of a supercluster is the most dense region, made up of a concentration of rich galaxy clusters and deep dark-matter potential wells \citep{Merluzzi2015_ShaSS,Higuchi_2020,Mercurio2015,2024A&A...689A.332A}. The prominent core of a supercluster is defined as the gravitationally bound region surrounding the most massive cluster(s) within the supercluster, corresponding to the peak of the density field. Massive rich clusters are predominantly found at the nodes of cosmic filaments \citep[e.g.,][]{2024A&A...681A..91E}. It is therefore natural to expect that these nodes, or at least some of them, are the cores of superclusters, so massive systems within superclusters can serve as markers of these cores. Although superclusters typically contain a single core, they can host more than three cores in various dynamical states \citep{2024PASA...41...78Z}. Recently, \citet{Sankhyayan2023} published an extensive catalogue of 662 superclusters identified using a modified FoF algorithm applied to the WHL cluster catalogue \citep{2015ApJ...807..178W}. Spectroscopic redshifts for 89$\%$ of the groups and clusters in the WHL catalogue were taken from SDSS DR12. This catalogue covers a redshift range $0.05 \leq z \leq 0.42$ and is complete\footnote{Completeness is defined as the detection rate of injected mock clusters using the cluster-detection algorithm, which reaches 100\% for clusters with a virial mass M${200c} > 2\times10^{14}$ M${\odot}$.}. Each supercluster in this catalogue has at least 10 member clusters.

Recently, Parekh et al. (in prep.) are studying the core of the superclusters from the sample of \cite{Sankhyayan2023} with VLA Sky Survey (VLASS) \citep{2020PASP..132c5001L}. VLASS is a multi-epoch S-band (2--4 GHz) all sky survey (dec $> -40^{\circ}$) with VLA B-configuration. The observation of all three epochs has been completed (the observation of the final 4$^{th}$ epoch is ongoing), and providing high-quality data products is undergoing. VLASS provides a fantastic angular resolution of $\sim$ 2.5$''$ necessary to mitigate blending effects and separate cores and lobes of radio galaxies. The 1$\sigma$ rms sensitivity of the single epoch image is $\sim$ 120 $\mu$Jy beam$^{-1}$. When data from all three epochs are combined, the sensitivity will improve to the $\sim$ 69 $\mu$ Jy beam$^{-1}$. From our initial inspection, we found a total of 75 ($\sim$ 11\%) radio BCGs in the VLASS data corresponding to the optical BCGs of the superclusters. We measured and compared the radio properties of BCGs with the global supercluster properties. In our analysis, we did not find radio-loud BCGs.  

\par Future high-resolution (sub-arcsec) and sensitive (sub-$\mu$Jy) SKA observations will significantly improve current statistics and image the central radio sources in superclusters with unprecedented detail. The SKA will greatly enhance the studies of BCGs across a wide range of spatial and spectral scales. SKA-Low (50--350 MHz) will be particularly powerful for detecting very steep-spectrum emission from aged radio plasma and cluster-scale diffuse components (haloes and mini-haloes) associated with BCG feedback, as well as probing the low-frequency curvature of their synchrotron spectra. The SKA-Mid Bands~1 and~2 (0.35--1.76 GHz) will trace most of the extended jets and lobes, providing high-fidelity total-intensity and polarization maps. At higher frequencies, SKA-Mid Band~5 (4.6--15.3 GHz) will isolate the compact AGN core, reduce the impact of synchrotron ageing and Faraday depolarization, and enable detailed studies of jet collimation, nuclear variability, and small-scale feedback process near to the BCG nucleus. Together, these SKA bands will allow spatially resolved, broadband spectral and polarimetric diagnostics of BCG radio emission, tightly linking nuclear activity to the thermodynamic state of the surrounding ICM and the larger-scale supercluster environment.

\par Therefore, it is important to study how central BCGs shape and regulate the cores of superclusters, since this in turn informs the global properties of the parent clusters and superclusters. Radio-loud BCG also inject seed particles for the production of large-scale diffuse radio sources, so understanding how they deposit energy into the surrounding medium is crucial to characterizing feedback at the heart of supercluster nodes. In the case of radio-loud BCGs, radio jets from AGN interacting with filamentary gas can highlight where and how baryons are funneled into supercluster nodes, providing an observational handle on the cosmic web–cluster connection. Furthermore, by comparing the radio luminosity and jet power of AGN with host cluster mass, we can test how the AGN feedback scales from clusters to superclusters, thus constraining the models of structure formation in $\Lambda$CDM cosmology.

\subsection{GRGs in clusters and superclusters \& the magnetic field of superclusters}

Giant Radio Galaxies (GRGs) represent the most extended class of radio galaxies, exhibiting projected linear sizes of $\geq 700$ kpc \citep{Dabhade2023}. Powered by active galactic nuclei hosting supermassive black holes, these systems launch relativistic jets that can reach dimensions comparable to, or exceeding, those of galaxy clusters. The recent discovery of a $\sim 7$ Mpc GRG by \citet{Oei2024_7Mpc} marks the largest known GRG to date. Owing to their extremely large extents, GRGs are likely influenced by the LSS of the cosmic web, particularly the overdense environments of galaxy clusters and superclusters \citep{Sankhyayan2024}. Understanding how these environments affect GRG formation and evolution provides important insights not only into the mechanisms that enable their exceptional growth, but also into the physics of the surrounding medium.

\citet{Sankhyayan2024} investigated the environmental dependence of GRG properties and found that $\sim$24\% of GRGs reside within groups or clusters, consistent with \citet{Oei2024_GRGs}, while only $\sim$5\% occupy supercluster regions. These results indicate a preference for GRGs to inhabit relatively sparse cosmic web environments. Nevertheless, the ambient density significantly influences their morphology: GRGs outside clusters tend to be larger than those within clusters, with a statistically significant ($>5\sigma$) size difference of $\sim$100 kpc. A similar pattern emerges when comparing GRGs in supercluster and non-supercluster environments, suggesting that large-scale overdensities can suppress the expansion of radio jets, a secondary yet measurable effect. Moreover, the environmental distribution shows that while only $\sim$21\% of non-supercluster GRGs are associated with clusters, as many as $\sim$77\% of GRGs within superclusters lie in cluster regions. This implies that the growth of GRGs in dense supercluster environments requires a sustained supply of energy and matter, typically provided by massive cluster reservoirs, to overcome the confining pressure of the ambient medium.

Beyond their role as tracers of AGN and jet evolution, GRGs also serve as valuable probes of intergalactic magnetic fields. Their high fractional polarization at low radio frequencies enables estimates of magnetic field strengths in superclusters via Faraday rotation measurements. When observed behind a supercluster, GRGs provide direct constraints on magnetic fields at levels of a few tens of nanogauss. Accounting for biases from the GRG host, intervening matter, and the Milky Way, \citet{Sankhyayan2024} identified eight GRGs behind six superclusters and estimated a magnetic field strength of $\sim$60 nG. \citet{Pignataro2025}, using a larger sample of background polarized sources behind three superclusters, obtained an estimate of $19^{+50}_{-8}$ nG, implying that the purely adiabatic compression of a primordial magnetic field is not sufficient to explain these observed scales. These studies highlight the potential of SKA to transform our understanding of magnetic field structures within the cosmic web and their interplay with the evolution of giant radio galaxies. With an expected 60–90 polarized sources per square degree across $\sim$30,000 square degrees in SKA1-MID Band 2 \citep{Heald2020}, the SKA is poised to profoundly advance the field.

\subsection{Galaxy populations in voids}
\label{sec_voids}
Cosmic voids are vast regions of space (sizes varying between $\sim$ 30 to $\sim$ 150 Mpc) that are nearly empty or a low matter of density \citep{1997A&AS..123..119E, Einasto2011_void_structure}. Typical small-size voids are associated with superclusters and located around the filaments. In contrast, some voids are giant over a hundred megaparsec across- separating the superclusters. With advancements in optical imaging and mapping, it is possible to identify voids within and around the supercluster region. Recently, a number of cosmic void catalogues have been made with optical data from the SDSS \citep{Sutter2012_void_catalog, Nadathur2016_boss_voids}. Most of the catalogues have low-z galaxy samples, which they used to compare the stellar mass distribution of galaxies inside and outside of voids. Observation shows that voids are not completely empty, but they contain very few galaxies compared to filaments and cluster nodes; hosting some dim galaxies, faint gas, or even galaxy groups. The specific populations depend on the large-scale environment of the void and the method used to detect it. Although most galaxies exist in the sheets between voids, the void population is vital for understanding large-scale structure formation and galaxy evolution in under dense environments. Furthermore, this peculiar galaxy population reside in a special environment that is undisturbed by the complex gravitational processes altering galaxies in massive clusters and groups. This population also provides a unique opportunity to investigate theories of cosmological structure formation. Compared to the dense clusters and filaments, very few voids have been observed and studied in radio wavelength. 

\citet{2011AJ....141....4K, 2012AJ....144...16K} have conducted H\,{\sc i} imaging of total 60 void galaxies in a nearby universe as part of the void galaxy survey (VGS) with Westerbork Synthesis Radio Telescope (WSRT). From this observation, they detected H\,{\sc i} emission in 41 galaxies, with masses ranging from $1.7 \times 10^{8}$ to $5.5 \times 10^{9}\,M_{\odot}$. The majority of these void galaxies are gas-rich, blue, low-luminosity disk or irregular systems, with stellar masses below $3 \times 10^{10}\,M_{\odot}$, and several show morphological or kinematic signatures of ongoing gas accretion, indicating that the void galaxies are still in the process of assembling. Interestingly, on small scales ($<600$ kpc) void galaxies exhibit clustering similar to that seen in higher-density environments, and 18 H\,{\sc i}-rich companions were identified. Contrary to predictions from simulations, no large population of faint, H\,{\sc i}-rich galaxies filling the voids was found. This suggests that void galaxies, while residing in extreme underdensities, are not fundamentally distinct from galaxies of similar luminosity and morphology in denser regions, but may still be assembling through continued gas accretion. 

\par The future high-sensitive SKA observations will significantly advance our understanding of void galaxies. With its unprecedented sensitivity (few $\mu$Jy beam$^{-1}$), the SKA will be capable of detecting extremely faint and low-mass H\,{\sc i}-rich dwarf galaxies ($M_{\rm HI} \sim 10^{6-7}\,M_{\odot}$), potentially revealing the predicted but as yet unseen population of galaxies inhabiting voids. Its ability to probe ultra-low column density gas ($<10^{18}\,\mathrm{cm}^{-2}$) will allow the detection of diffuse, extended reservoirs of neutral hydrogen and possible filamentary inflows connecting galaxies to the intergalactic medium. High-resolution kinematic studies will clarify the role of cold gas accretion and ongoing assembly in isolated void environments. Moreover, the SKA will detect H\,{\sc i}-rich galaxies up to $z \sim 1$, yielding statistical samples of millions of galaxies across a wide range of environments, allowing evolutionary studies of void galaxies over cosmic time. Finally, by mapping void galaxy distributions and dynamics, the SKA will also offer powerful constraints on cosmology, dark energy, and alternative theories of gravity.

\section{Summary and conclusions}

On the largest scales, the Universe exhibits a web-like structure composed of massive filaments separated by giant voids. Cosmological simulations have revealed the detailed three-dimensional and dynamical evolution of this network. Continuous accretion of matter along filaments, together with the hierarchical merging of smaller substructures, leads to the formation of gravitationally bound galaxy clusters at the nodes and, on even larger scales, superclusters. Tenuous baryonic plasma is drawn into the deep gravitational potentials of these structures. While large optical surveys trace the galaxy distribution and thereby map the underlying dark-matter skeleton, the diffuse baryonic plasma residing throughout the cosmic web - accessible only through faint X-ray, Sunyaev–Zel’dovich, and synchrotron radio signals - has yet to be conclusively detected.

The gravitational energy released during the evolution of the LSS is dissipated through shocks and gas motions across the whole cosmic web heating up the intergalactic medium and the ICM in clusters. A small fraction of this energy is also employed in the acceleration of cosmic rays and the amplification of magnetic fields. The interaction between such CRs and the intergalactic magnetic fields produces different levels of synchrotron radiation, providing us with a unique probe to study the processes dissipating the gravitational energy derived from structure formation. Many simulations predict and investigate this synchrotron
emission \citep[e.g.][]{Dolag2008SSRv,donnert09,pfrommer06,Vazza2015A&A, vazza19,boss24}, but so far its direct detection has been achieved only in galaxy clusters (in the form of radio halos and radio relics) and, very recently in a few cases outside clusters of galaxies.

These discoveries beyond the scale of individual galaxy clusters have been obtained thanks to the advent of the new generation of radio telescopes, such as LOFAR, the uGMRT, MeerKAT. Radio Megahalos, filling the whole volume of a few clusters have been recently discovered (Section \ref{Sec:megahalos}). Their origin is still unclear and the available statistics is still scarce, owing to their extreme low surface brightness, but the current data suggest that megahalos might be a common phenomenon that could be unveiled with deeper observations, such as those provided by SKA.
Also recently, radio bridges of diffuse, faint synchrotron emission have been detected in the region between a few cluster pairs (Section \ref{sec:bridges}). Their properties, combined with tailored cosmological simulations suggest that turbulence plays a crucial role in accelerating particles already in the pre-merger state, but increasing the number of known radio bridges is necessary to confirm this scenario.
Bridges and Megahalos represent the unique direct detections of synchrotron emission beyond galaxy clusters so far. Evidence of the presence of cosmic rays and magnetic fields in cosmic filaments have been reported only through stacking experiments \citep{vernstrom21, vernstrom23, hou23}. 

In Section \ref{sec:simulations} we presented an MHD simulation aimed at predicting the radio emission from the cosmic web as it will be visible with SKA-Low and SKA-Mid in AA4 configuration. The simulation includes an initial seed magnetic field, additional magnetic fields released by AGNs and star formation feedback. The simulation tracks the propagation of cosmic rays injected by shocks, AGNs and star formation. The synchrotron emission of these particles is computed in post-processing as a function of frequency. Fig. \ref{fig:cosmo_sim} summarises our findings: thanks to it combination of sensitivity to the large scale emission and observing frequency, SKA-Low is better suited at detecting the synchrotron emission from cosmic filaments. In particular, SKA-Low will have the sensitivity to detect faint extended emission from the brightest regions of the cosmic web with 1 hour observation. SKA-Mid will achieve similar detection capabilities with 10 hours observation and point source subtraction with subsequent taper of the long baselines.  

We notice that complementary probe of the magnetisation of the cosmic web is the statistical analysis of Faraday Rotation from polarised background sources, whose theoretical predictions do not depend on the assumed distribution of cosmic ray electrons. The application of Faraday Rotation analysis in the SKA landscape is the subject of another Chapter in this volume \citep{OSullivan01.2026.SKA}.

Radio galaxies, in clusters, superclusters and cosmic voids can also be used to study the cosmic web (Section \ref{sec: BCG}). They are relevant through the whole LSS because they are a reservoir of cosmic rays and can be used to measure magnetic fields. BGCs are particularly important because they shape and regulate the cores of clusters and superclusters. GRG give us information about the medium they propagate through. Furthermore, SKA will allow us to perform studies of galaxies in voids providing strong constraints for cosmology.

The advent of SKA promises the identification of many new diffuse radio sources in galaxy clusters and beyond on different scales. Conventional manual cataloguing methodologies will be insufficient to exploit the capabilities of radio Surveys performed with SKA. Convolutional neural networks, trained on synthetic radio observations built upon cosmological simulations and designed to detect faint extended sources in radio surveys have been proposed \citep{stuardi2024, sanvitale25}. Generative AI, such as diffusion models or GANs, can be a further step forward. In fact, there have been recent attempts to generate realistic synthetic radio maps that match observed properties \citep{vicanek}, including features such as radio haloes, relics, or AGN jets. This helps in synthesizing high-fidelity images that mimic actual radio morphology and spectrum characteristics. These synthetic datasets can augment training sets for other machine learning pipelines and improve prediction accuracy for unseen cluster observations.

We propose below a list of key observations that should be performed with the SKA telescopes in order to explore the LSS of the Universe with unprecedented detail:
\begin{itemize}
    \item SKA-Low observations of the most massive clusters in the southern sky at low-mid redshift (z$\sim 0.05-0.2$) to search for megahalos in the most promising targets but also possible accretions shocks and the filamentary entry at end of filaments. Complementary SKA-Mid observations of the same targets would give us spectral and polarization information, crucial to constrain the models.
    \item Deep observations of the known bridges (A1758 is too far north but the other ones mentioned in Section \ref{sec:bridges} can be observed with SKA). The sensitivity of the SKA, both Low and Mid, would allow us to image these bridges at relatively high resolution ($\sim 10-20$ arcsec), define their morphology and spectral index distribution, crucial to understand their origin.
    \item In order to search for new radio bridges, we could follow up the X-ray detected bridges with eROSITA in the Southern Sky. A clear example is the 13 Mpc long X-ray bridge between A3667 and A3651 \citep{dietl24}.
    \item Exploit the forthcoming optical surveys to algorithmically identify the densest regions of the cosmic web (see Section~\ref{algo-structure}) and observe them with SKA to search for synchrotron emission. Cosmic void will also be identified in the same Surveys and could be targets to search for the faint population of radio galaxies expected in the less dense regions of the Universe. In particular, the Rubin Observatory Legacy Survey of Space and Time (LSST) will deliver deep, uniform, multi-band optical imaging over roughly 18,000 deg$^{2}$ of the Southern Sky, producing extremely large photometric galaxy samples that underpin precision measurements of how galaxies trace the matter distribution \citep{Ivezic2019LSST}. Complementing this, the 4-metre Multi-Object Spectroscopic Telescope (4MOST) will obtain optical spectra and deliver redshifts for tens of millions of objects ($>$25 million in its first five years), providing the accurate distance information and environmental context needed to map the three-dimensional large-scale structure with high fidelity \citep{Verdier2026_4MOSTCRS}. 
   
\end{itemize}

\section*{Acknowledgements}

The authors would like to thank the referee for the suggestions which improved the presentation of this chapter. FV acknowledges the CINECA award  "IscrB\_CREW"  under the ISCRA initiative, for the availability of high-performance computing resources and support. SP acknowledges Manipal Centre for Natural Sciences, Centre of Excellence, Manipal Academy of Higher Education (MAHE) for facilities and support. SS acknowledges the support from the Estonian Research Council grant PSG1045. AB acknowledges support from the ERC CoG $\vec{B}ELOVED$, GA n. 101169773. FV acknowledges funding under the European Union’s  Horizon Europ rogram through the ERC Synergy Grant COSMOMAG (Project Id. 101224803).

\bibliographystyle{abbrvnat-maxbibnames4}
\bibliography{chapter}

\end{document}